
\magnification=1200
\vskip 0.5cm
\def\dsl{\raise.15ex\hbox{/}\kern-.57em\partial}
\def\Dsl{\,\raise.15ex\hbox{/}\mkern-13.5mu D} 
\def\Asl{\,\raise.15ex\hbox{/}\mkern-13.5mu A} 
\def\Bsl{\,\raise.15ex\hbox{/}\mkern-13.5mu B} 

\def\L{{\cal L}}
\def\bP{{{\bar \psi}^r}}
\def\nP{{{\psi}^r}}
\def\dL{{\delta L}}
\def\dd{{\delta \partial_+}}
\def\Ch{{\chi^r}}
\def\bCh{{{\bar \chi^r}}}

\def\dl{{\delta (x_- - x^{'}_-)}}
\def\ab{{\delta^{ab}}}
\def\rs{{\delta^{rr^{'}}}}
\def\pP{{{\bar \psi}^{r^{'}}}}
\def\D{{\cal D}}
\font\title=cmr12 at 12pt
\footline={\ifnum\pageno=1\hfill\else\hfill\rm\folio\hfill\fi}
\baselineskip=18pt
\vskip 1.0cm
\centerline{\title PATH INTEGRAL APPROACH TO TWO-DIMENSIONAL QCD}
\centerline{\title IN THE LIGHT-FRONT}
\vskip 1.5cm
\centerline{{\bf P. GAETE}}
\centerline{\it Instituto de F\'{\i}sica, Universidade
Federal do Rio de Janeiro}
\centerline{\it C.P 68528, BR-21945, R.J., Brazil}
\vskip 0.10cm
\centerline{{\bf J. GAMBOA}\footnote{$^{\dagger}$}{Supported by Alexander
von Humboldt Foundation.}}
\centerline{\it Fachbereich 7 Physik, Universit\"at Siegen, Siegen}
\centerline{{\it D-57012, Germany}}
\vskip 0.1cm
\centerline{{\bf I. SCHMIDT}\footnote*{Work supported in part
by FONDECYT (Chile), contract 1931120.}}
\centerline{\it Departamento de F\'{\i}sica, Universidad T\'ecnica Federico
Santa Maria}
\centerline{\it Casilla 110-V, Valparaiso, Chile}
\vskip 1.5cm
{\bf Abstract}. Two-dimensional quantum chromodynamics in the light-front frame
is studied following hamiltonian methods. The theory is quantized using the
path integral formalism and an effective theory similar to the
Nambu-Jona Lasinio model is obtained. Confinement in two
dimensions is derived by analyzing directly the constraints in the
path integral.
\vskip 1.2cm
\leftline{Si-93-10}
\leftline{USM-TH-63}
\leftline{IF/UFRJ/93/11}
\leftline{October 1993}
\vfill \eject
\centerline{\bf 1. Introduction}
\bigskip
Quantum field theory quantized in the light-front has been extensively
studied in the past few years as an alternative way for understanding
non-perturbative phenomena
{\bf [1]}. Although this approach is quite old {\bf [2]}, only recently new
techniques of calculation have been developed {\bf [3, 4]} that
could allow, in principle, the study of phenomena such as
confinement or hadronization, which are very difficult to
understand through the conventional approach.

Several years ago, t'Hooft studied the solubility of
QCD in two dimensions $(QCD_2)$ in the light-front frame, introducing the
${1\over N}$ expansion {\bf [5]}. In his work he was able to
solve the theory
in the large $N$ limit, and then show how the bound state spectrum can be
obtained by solving a Bethe-Salpeter equation in this limit.
However, in spite of the relevance that the t'Hooft results could have
in our understanding of QCD in four dimensions, not much further
progress was reached at that time
in order to understand the perturbative and non-perturbative structure of QCD
in the light-front frame.

The revival of light-front quantization has been mainly
pioneered by the authors of reference {\bf [3]}, and later
also by those of reference {\bf [4]}. In these references
two different non-perturbative methods for calculating
light-front wave functions have been proposed. These wave
functions are the amplitudes for a Fock state expansion of the
hadronic bound states, in a Hamiltonian light-front approach to QCD.
Both methods although promising still present some technical
difficulties despite intense recent research {\bf [1]}.

The canonical structure of QCD in the light-front frame and his
subsequent quantization via the path integral method is to our
knowledge still an open problem.\footnote{$^1$}
{In reference {\bf [6]} the Hamiltonian
formulation of the Schwinger model is studied at the classical
level. In this sense, this reference is a particular case
of our results. On the other hand, in reference {\bf [7]} the
reader can find recent canonical studies of two dimensional
light-front gauge theories.} In the past, this approach has
been very useful in the understanding of many aspects of gauge theories
and we will show that it is very useful in the present context
as well.

The purpose of this research is the study of the
canonical structure of light-front $QCD_2$ and its quantization
following the path integral method. Our main results are: (a) we
find the complete canonical structure (constraints and algebra),
(b) the effective field theory obtained after the
integration of the gauge field is derived from first principles,
and (c) we justify some results obtained by other authors using
different methods.

The article is organized as follows: in section 2 the hamiltonian
formulation is carried out. In section 3 we discuss the problems
asociated with gauge fixing and we quantize the theory following
the path integral method. Finally in section 4 we present some conclusions.

\vfill \eject
\centerline{\bf 2. $QCD_2$ in Light-Front Coodinates: Hamiltonian}
\centerline{\bf Analysis}

\bigskip
In this section we study the canonical structure of $QCD_2$ in light-front
coordinates following the Dirac constrained theory {\bf [8]}.

Let us start by considering the lagrangian density
$$\L = -{1\over 4}F^a_{\mu \nu} F^{a\mu \nu} + \bP (i \Dsl -
m)\nP, \eqno(2.1)$$
$(\mu = 0,1; a = 1,2,..., N^2; r = 1,2,..., N)$ \hfill \break
where the sum convention is assumed. Here $\nP$ are the $r$ quarks fields
and the strenght tensor $F^a_{\mu \nu}$ and the covariant derivative
are defined as
$$\eqalignno{& F^a_{\mu \nu} = \partial_\mu A^a_\nu - \partial_\nu A^a_\mu
- g f^{abc} A^b_\mu A^c_\nu, &(2.2)
\cr &
D_\mu = \partial_\mu + ig A^a_\mu T^a, &(2.3) \cr}$$
where $g$ is the coupling constant and $T^a$ and $f^{abc}$ are
the generators and structure constants associated with the gauge
group $SU(N)$.

In the light-front frame approach one defines the coordinates
$$x^\pm = {1\over \sqrt{2}} ( x^0 \pm x^1), \eqno(2.4)$$
and then  writes all the quantities involved in the lagrangian (2.1)
in terms of $x^\pm$ instead of $x^0$ and $x^1$.
After doing this, the lagrangian density (2.1) becomes
$$\L = {1\over 2}{(F^a_{+-})}^2 + \bP ( i \gamma_- D_+ +
i \gamma_+D_- - m)\nP , \eqno(2.5)$$
where $\gamma_\pm$ and $D_\pm$ are defined in complete analogy
with (2.4), {\it i.e}
$$\eqalignno{& \gamma_\pm = {1\over \sqrt{2}}(\gamma_0 \pm \gamma_1),
\cr &
D_\pm = {1\over \sqrt{2}} (D_0 \pm D_1), &(2.6) \cr}$$
and the $\gamma_\pm$ matrices satisfy
$$\gamma^2_\pm = 0,\quad \{\gamma_+, \gamma_-\} = 2. \eqno(2.7)$$

In order to carry out the hamiltonian formulation, we are forced to choose
a time coordinate, which is usually chosen as $x^+$. Thus the canonical
momenta are
$$\eqalignno{& \pi^a_+ = {\dL\over \dd A^a_+} = 0, &(2.8a)
\cr &
\pi^a_- = {\dL\over \dd A^a_-} = F^a_{+-}, &(2.8b)
\cr &
P^r = {\dL \over \dd \nP} =  i \bP \gamma_-, &(2.8c)
\cr &
{\bar P}^r = {\dL\over \dd \bP} = 0, &(2.8d) \cr}$$
where $L = \int dx_- \L$.

Observing (2.8) one can see that there are three primary
constraints, namely
$$\eqalignno{& \pi^a_+ = 0, &(2.9a)
\cr &
\chi^r = P^r -  i \bP \gamma_- = 0, &(2.9b)
\cr &
{\bar \chi^r} = {\bar P^r} = 0, &(2.9c), \cr}$$
and which must be preserved in time $x^+$.

The total hamiltonian can then be computed, with the result
$$\eqalignno{H_T = & \int dx_- \biggl[ {1\over 2} {(\pi^a_-)}^2 + \pi^a_-
\partial_- A^a_+ + g \pi^a_- f^{abc} A^b_- A^c_+
+ g\bP \gamma_- \nP A^a_+ T^a
\cr &
- i \bP \gamma_+ D_- \nP - m \bP \nP +
u^a_0 \pi^a_+ + u^r_1 \chi^r + u^r_2 {\bar \chi^r} \biggr], &(2.10) \cr}$$
where $u^a_0, u^r_1$ and $u^r_2$ are Lagrange multipliers.

One can see that the preservation in time of $\pi^a_+$ implies the
secondary constraint
$$G^a = \partial_- \pi^a_- + g \pi^c_- f^{abc} A^b_- + g\bP
\gamma_- \nP T^a , \eqno(2.11)$$
and that the other constraints $(\chi^r, {\bar \chi^r})$ do not generate new
constraints.

A straightforward analysis shows that the constraints $G^a, \Ch$ and $\bCh$
are second class while $\pi^a_+$ is first class. On the other hand, a simple
inspection shows also that $(G^a, \Ch, \bCh)$ are not a minimal number of
second class constraints. The minimal set is found by combining appropiately
$G^a, \Ch$ and $\bCh$, and it is straightforward to verify that this set is
$$\eqalignno{&\Omega^a_0 = \pi^a_+, &(2.12a)
\cr &
\Omega^a_1 = G^a + i ({\bCh} T^a \nP + {\bP} T^a {\Ch}) , &(2.12b)
\cr &
\Ch = P^r - i\bP \gamma_-, &(2.12c)
\cr &
\bCh = {\bar P^r}, &(2.12d) \cr}$$
where $(\Omega^a_0, \Omega^a_1)$ and $(\Ch, \bCh)$ are first and second class
constraints respectively.

The first class constraints satisfy the algebra
$$\left\{ \Omega^a_\mu (x_-), \Omega^b_\nu (x^{'}_-)\right\} = 0, \eqno(2.13)$$
while the non-vanishing Poisson bracket between $\Ch$ and $\bCh$ are
$$\left\{ \Ch_\alpha (x_-) , {\bar \chi}^{r^{'}}_\beta (x^{'}_-) \right\} =
- i{(\gamma_-)}_{\beta \alpha} \delta^{r r^{'}} \delta (x_- - x^{'}_-),
\eqno(2.14)$$
where $\alpha, \beta$ are spinorial indices.

In order to eliminate the second class constraints we define the usual Dirac
bracket. In this case the non-vanishing Dirac brackets between the canonical
variables are
$$\eqalignno{& \{ A^a_+ (x_-), \pi^b_+ (x^{'}_-) \}_{DB} = \dl \ab=
\{ A^a_- (x_-), \pi^b_- (x^{'}_-) \}_{DB}, &(2.15a)
\cr &
\{ \nP_\alpha (x_-), \pP_\beta (x_-^{'})\}_{DB} =
{(\gamma_-)}_{\beta \alpha}^{-1} \rs \dl, &(2.15b)
\cr &
\{\nP_\alpha (x_-), P^{r^{'}}_\beta (x^{'}_-) \}_{DB} = \rs \ab \dl, &(2.15c)
\cr &
\{ \bP_\alpha (x_-), {\bar P}^{r^{'}}_\beta (x^{'}_- ) \}_{DB} = \rs \ab \dl.
&(2.15d) \cr}$$

The set of equations (2.10), (2.12-2.15) defines completely the canonical
structure of the theory. The next step is to fix the gauge in order to
quantize the theory.

\vfill \eject

\centerline{\bf 3. Gauge Fixing and the Path Integral Quantization}

\bigskip
In this section we discuss the quantization of the previous model following
the path integral approach. There are several reasons that justify this study:
(i) to our knowledge the quantization of $QCD_2$ in
the light-front frame following the path integral approach
has never been discussed before; (ii) this
study could throw some light into the derivation of the Feynman rules in the
light-front quantization method and the influence of the
zero modes; (iii) the path integral approach could allow for the
inclusion of new fields that could simplify the perturbative
structure of the theory.

With these facts in mind, in this section we try to clarify the problem
of gauge fixing and path integral quantization of $QCD_2$ in
the light-front frame.
\vskip 0.25cm
$\underline {{\it Gauge\,\,\, Fixing}}$

The gauge freedom is reflected from the hamiltonian point of view in the
presence of first class constraints. In the problem at hand, we have
two first class constraints (eqs. (2.12a-b)) and, as a
consequence, two conditions are necessary in order to
fix completely the gauge freedom.
Thus, we can start by imposing the following condition as gauge fixing
$$\Omega^a_2 = A^a_- = 0, \eqno(3.1)$$
which is known as light-cone gauge and which must be imposed as a new
constraint of the theory. Following Dirac's method {\bf [8]}, (3.1) must
be preserved in time, {\it i.e.}
$$\partial_+ \Omega^a_2 = \{ \Omega^a_2 , H_T\} = 0. \eqno(3.2)$$

Computing (3.2), we find that this consistency condition implies the new
constraint
$$\Omega^a_3 = \pi^a_- + \partial_- A^a_+ + g f^{abc} A^b_-A^c_+
= 0. \eqno(3.3)$$

The conditions (3.1) and (3.3) fix completely the gauge freedom.
In fact, computing the Poisson algebra we find that the
non-vanishing brackets between the first
class constraints and the gauge conditions are
$$\{ \Omega^a_0 (x_-), \Omega^b_3 (x^{'}_-) \} = ( \ab
\partial_- - g f^{abc}A^c_- )\dl, \eqno(3.4a)$$
$$\eqalignno{\{ \Omega^a_1 (x_-), & \Omega^b_3 (x^{'}_-) \} =
 g f^{abc} (A^c_+ \partial_-  \dl  + \partial_- A^c_+ \dl )
\cr &
+ g f^{abc} \pi^c_-  \dl
+ g^2 f^{afc}f^{bgc} A^f_- A^g_+  \dl, &(3.4b) \cr}$$
which is a second class constraint algebra.

The question now is: are there other alternative gauge fixing conditions
besides $A_-^a = 0$~?~. The answer to this
question is, of course, yes, although the correct way to
implement other possible gauge fixing conditions in the light
front-frame is not trivial.

One could try to find, for instance, the analogous of the gauge fixing in
a covariant gauge theory, but this
procedure does not work here. Indeed, this can be verified
by constructing the analogous of the
Lorentz gauge $\partial_\mu A^{a\,\mu} = 0$ in the light-front
$$\partial_-A^a_+ + \partial_+A^a_- = 0, \eqno(3.5)$$
but a simple analysis shows that this condition does not fix the gauge
freedom. In fact, it can be shown that (3.5) is not a true gauge condition
because when preservation in time is imposed, we cannot generate a new
constraint fixing the remaining gauge symmetry.

The same occurs when we consider the analogous of the axial gauge
$n^\mu A^a_{\mu} =0$ in the light-front
$$n_-A_+^a + n_+A_-^a = 0. \eqno(3.6)$$

A possible solution to this problem consists in modifying
slightly the previous gauge
conditions. Using (3.5) and (3.6) one can see that the unique
possible choices
for the above gauge conditions are\footnote{$^2$}{The other
possible choices
$\partial_- A^a_+ = 0$ for the Lorentz gauge and $n_- A^a_+ = 0$ for the
axial gauge do not fix the gauge freedom.}
$$\partial_+ A^a_- = 0 \,\,\,\,(Lorentz\,\, gauge), \eqno(3.7)$$
$$n_+ A^a_- = 0 \,\,\,\,(Axial\,\,gauge).\eqno(3.8)$$

The hamiltonian analysis shows, however, that equations (3.7) and
(3.8), although they are two independent gauge conditions that fix completely
the gauge freedom, are contained in (3.1). This last point
can be shown by computing the analogous of (3.2) using (3.7) and (3.8).
This calculation gives
$$\partial_+ \Omega^a_3 = 0 = n_+ \Omega^a_3, $$
and, as a consequence, to impose (3.7) or (3.8) is formally the same
condition (3.1).

This last result means that in two dimensions in the light-front
frame, the light-cone gauge (3.1) contains a complete family of
gauge conditions that simplify the canonical
analysis.\footnote{$^3$} {Recently there has been some discussion
about the problem of fixing the residual gauge in
light-front $QCD_4$ {\bf [9]}. In this paper we
assume periodic boundary conditions, in which case the zero modes
do not have to be considered.} This last
point is another advantage of the light-front approach.

\vskip 0.25cm
$\underline{{\it Path\,\,\, Integral\,\,\, Quantization}}$

The quantization in this case must be performed using the modication
introduced by Senjanovic {\bf [10]} because there are second
class constraints.

The generating functional is
$$ \eqalignno{ Z = &\int \D \pi^a_+ \D A^a_+ \D \pi^a_- \D A^a_- \D \bP \D \nP
\D P^r \D {\bar P}^r \times
\cr &
\det \| M^{ab} \|\,\,
{ det \| \{\chi_\alpha, \chi_\beta\}\|}^{1\over 2} \,\,\delta (\Omega^a_0)
\,\,\delta ( \Omega^a_1) \,\,\delta (\Omega^a_2) \,\,\delta (\Omega^a_3)\,\,
\delta (\chi^r)\,\,
\delta ({\bar \chi^r})
\cr &
\exp\biggl[ i\int dx_+ dx_- \biggl( \pi^a_+ \,\partial_+ A^a_+ +
\pi^a_- \,\partial_+ A^a_- + P^r \partial_+ \nP + {\bar P}^{r} \partial_+\bP
\cr &  - {1\over 2}{(\pi^a_-)}^2 - \pi^a_-\partial_-A^a_+
- g\pi^a_- f^{abc}
A^b_- A^c_+ - g\bP \gamma_- T^a \nP A^a_+
\cr & +  i\bP \gamma_+
D_- \nP - m \bP \nP \biggr) \biggr]. &(3.9) \cr}$$

Using (3.4), the first determinant can be explicitly computed
$$\det \,\|M^{ab}\|\, = \det \left(\matrix{0&\kappa^{ab}\cr
-\kappa^{ab}& \rho^{ab}\cr}\right) \dl, \eqno(3.10)$$
where its elements are
$$\eqalignno{& \kappa^{ab} = \ab \partial_- - gf^{abc}A^c_-,
\cr &
\rho^{ab} = gf^{abc}\partial_-A^c_+ +gf^{abc}A^c_+\partial_- +gf^{abc}\pi^c_- +
g^2 f^{afc}f^{bgc} A^f_+ A^g_-, \cr}$$
while that the second determinant
$$\det \{ \Ch_\alpha, {\bar \chi}^{r^{'}}_\beta \} = \det \left(\matrix {
0&-i{(\gamma_-)}^{-1}_{\beta \alpha}\cr -
i{(\gamma_-)}_{\alpha \beta}&0\cr}\right)
\dl \rs, \eqno(3.11)$$
is a c-number and can be dropped off the path integral as a
normalization factor.

Integrating in $\pi^a_-, \pi^a_+, A^a_-, {\bar P}^r, P^r$ and exponentiating
(3.10) in terms of anticommutative ghosts, we find
$$\eqalignno{Z = &{\cal N}\int  \D A^a_+ \D \bP \D \nP \D \,\,(ghosts)
\,\,\delta [-\partial^2_- A^a_+ + g \bP \gamma_- T^a \nP ]\times
\cr &
\exp \biggl[ i\int dx_+ dx_- \biggl( {1\over 2}{(\partial_- A^a_+)}^2
+ i\bP \gamma_+ \partial_- \nP +  i\bP \gamma_- \partial_+ \nP
\cr &
- g\bP \gamma_- T^a\nP A^a_+ - m\bP \nP + (ghosts) \biggr) \biggr],
&(3.12) \cr}$$
where ${\cal N}$ is a normalization factor.

Using the antisymmetry of the structure constants and the anticommutative
character of the ghosts, it is easy to see that there is no coupling between
the ghosts and the gauge fields and that the integration in
the ghosts fields is
trivial. This result is a consequence of a general statement that establishes
that all axial like gauges are free of interactions with the
ghosts {\bf [11]}.
Having this fact in mind, (3.12) becomes
$$\eqalignno{Z = &{\cal N}^{'} \int \D A^a_+ \D \bP \D \nP \,\,
\delta [ -\partial^2_- A^a_+ + g\bP \gamma_- T^a \nP ]\times
\cr &
\exp \biggl[ i \int dx_+ dx_- \biggl( {1\over 2}{(\partial_- A^a_+)}^2
+  i\bP \gamma_+ \partial_- \nP +  i\bP \gamma_- \partial_+ \nP
\cr &
-g \bP \gamma_- T^a \nP A^a_+ - m \bP \nP \biggr) \biggr].  &(3.13)\cr}$$
This formula gives the path integral version of ${QCD}_2$.

The constraint (Gauss's law):
$$\partial_-^2A_+^a - J_-^a = 0,  \eqno(3.14)$$
with $J_-^a = g\bP \gamma_- T^a \nP $, physically can be understood
as follows. In two dimensions there only exists an electric
field, which is given by $E^a = \partial_-A_+^a$. Using this
fact and equation (3.14), $E^a$ is given by
$$E^a(x_-) =  \int dx_-^{'} \epsilon (x_--x_-^{'})
J_-^a(x_-^{'}), \eqno(3.15)$$
where $\epsilon (x_--x_-^{'})$ is the sign function.
In order to see what this result means, let us assume for the
moment that the quarks are point-like. Thus $J_-^a(x_-)$ can be
written as
$$J^a_-(x_-) = \sum_{\alpha = 1}^N q_{\alpha}\,\delta
(x_--x_-^{\alpha})T^a, \eqno(3.16)$$
where $q_{\alpha}$ is the quark charge and $x_-^{\alpha}$ is the
place where the $\alpha$-th quark is localized. Using (3.16),
the electric field becomes
$$E^a(x_-) =  \sum_{\alpha = 1}^N q_{\alpha}\,\epsilon
(x_--x_-^{\alpha})T^a . \eqno(3.17)$$
Therefore the electric field between the particles is unable to
spread out and the quarks are confined.
The reader should notice that confinement is
present irrespective of the (non)abelian character of the gauge
field {\bf [12]}.

The next step is to integrate the $A^a_+$ field. Using the identity
$$\int \D \phi \,\,\delta ( A\phi + B)\,\, e^{i\,\int dx (\phi\,C\,\phi +
D\, \phi )} = e^{i\,\int dx ( {B\over A} C {B\over A} - {D\over A} \,B)},
\eqno(3.18)$$
where $A,B,C,D$ are operators (with inverse),
equation (3.13) becomes
$$\eqalignno{Z = &{\cal N}\int \D \bP \D \nP
\exp \biggl[ i\int dx_+ dx_- \biggl( \bP
(i\gamma_+ \partial_- + i\gamma_- \partial_+ - m) \nP \cr
&-{1\over 4} \int dx_-^{'} J^a_-(x_-)|x_--x_-^{'}|J^a_-(x_-^{'})
 \biggr) \biggr],   &(3.19) \cr}$$
where $J^a_-(x_-) = g \bP (x_-)\gamma_- T^a \nP (x_-)$ and
$|x_--x_-^{'}|$ is the propagator of the gluon field obtained by
inverting the operator $\partial_-^2$.

We would like to add a few sentences about the procedure used
here for inverting the operator $\partial_-^2$. The central observation
consists in noticing that the invertion of $\partial_-^2$
is a similar mathematical problem to the calculation of the propagator of
a massless relativistic
particle {\bf [13]} moving in a one-dimensional space. Following this
analogy, ${1\over \partial_-^2}$ can be represented by the integral
$$\int_0^\infty dT\,\,T^{-{D\over 2}}\, \exp \biggl[-{{\Delta
{\vec x} }^2\over 2T}\biggr], \eqno(3.20)$$
where for generality, we are supposing a $D$-dimensional operator and
$\Delta {\vec x} = {\vec x }- {\vec x^{'}}$.

The integral (3.20), however, is infrared divergent and it is necessary
to perform a regularization for extracting a finite result. In order to
regularize, we replace (3.20) by
$$\int_0^\infty dT T^{-{D\over 2}} \exp \biggl[
-{({\Delta {\vec x})}^2\over 2T} -{m^2\over 2}T\biggr],
\eqno(3.21)$$
where $m^2$ is a massive regulator that is put equal to zero
at the end of the calculation.

The integral (3.21) can be computed explicitly and gives the
following result
$$\sim {\biggl[ {\Delta \vert {\vec x}\vert\over
m}\biggr]}^{{1\over 2} - {D\over  4}}
K_{{1\over 2} - {D\over 4}} (m \Delta \vert{\vec x}\vert), \eqno(3.22)$$
where $K_{{1\over 2} - {D\over 4}}$ is the Bessel function.

The massless limit corresponds to the replacement of the Bessel function by
its asymtotic behaviour when $m \rightarrow 0$, {\it i.e.} $\sim
{[ m \Delta \vert{\vec x}\vert ]}^{{1\over 2} - {D\over 4}}$.
Then the limit becomes regular and
we find for $D=1$ the result that appears in (3.19).

Equation (3.19) is the starting point for the perturbative and
non-perturbative evaluation of quantum corrections of $QCD_2$.
Non-perturbati-vely this formula was used in reference {\bf [14]}
in order to derive the t'Hooft equation for the bound states in
the large-N limit.

{}From the canonical point of view, the effective action that
appears in equation (3.19) was used by the authors of reference
{\bf [15]} for the numerical study of light-front quantized
$QCD_2$.

Finally, we should mention that the Higgs mechanism in
light-front quantized field theory was also considered in
reference {\bf [16]}.
\vskip 0.5cm
\centerline{\bf 4. Conclusions}

\bigskip
We summarize our results. In this paper we have studied the
canonical structure of $QCD_2$ in the light-front frame and have
also quantized the theory using the path integral formalism.

We have discussed some problems with the gauge fixing procedure
and we have shown
how the Lorentz or axial like gauges in the light-front are contained
in the light-cone gauge.

The path integral quantization was performed using the Senjanovic method
due to the existence of second class constraints and an effective field theory
for the fermionic fields that corresponds to a Nambu-Jona Lasinio model was
obtained. This effective field theory is the starting point for some approachs
to $QCD_2$.
\vfill \eject
\vskip 1cm
\centerline{\bf REFERENCES}
\bigskip
\item{{\bf [1]}} For a review and references, see S. J. Brodsky,
G. McCastor, H.-C. Pauli and S. S. Pinsky, {\it Particle World}
{\bf 3} (1993) 109
\item{{\bf [2]}} P. A. M. Dirac, {\it Rev. of Mod. Phys.} {\bf 21}
(1949) 392; S. Weinberg, {\it Phys. Rev.} {\bf 50} (1966) 1313.
\item{{\bf [3]}} H. C. Pauli and S. J. Brodsky, {\it Phys. Rev.} {\bf
D32} (1985) 1993; ibid. {\bf D32} (1985) 2001.
\item{{\bf [4]}} R. J. Perry, A. Harindranath and K. G. Wilson,
{\it Phys. Rev. Lett.} {\bf 65} (1990) 2959; D. Moustaki, S. Pinsky,
J. Shigenitsu and K. G. Wilson, {\it Phys. Rev.} {\bf D43} (1991) 3411.
\item{{\bf [5]}} G. t'Hooft, {\it Nucl. Phys.} {\bf B75} (1974) 461.
\item{{\bf [6]}} D. Moustaki, {\it Phys. Rev.} {\bf D42} (1990) 1184.
\item{{\bf [7]}} F. Lentz, M. Thies, S. Levit and K. Yazaki,
{\it Ann. of Phys.} {\bf 208} (1991) 1; T. Heinzl, S. Krusche
and E. Werner, {\it Phys. Lett.} {\bf B275} (1992) 410;
K. Hornbostel, {\it Phys. Rev.} {\bf D45} (1992) 3781.
\item{{\bf [8]}} See for example: K. Sundermeyer, {\it
Constrained Hamiltonian Systems}, Springer Verlag, 1984.
\item{{\bf [9]}} W.-M. Zhang and A. Harindranath, {\it Phys. Lett.}
{\bf B314} (1993) 223.
\item{{\bf [10]}} P. Senjanovic, {\it Ann. of Phys.} {\bf 100}
(1976) 227.
\item{{\bf [11]}} See for example: G. Leibbrandt, {\it Rev. of
Mod. Phys.} {\bf 59} (1987) 1067; W. Konetschny, {\it Phys. Lett.
}{\bf B90}
(1980)263.
\item{{\bf [12]}} S. Coleman, R. Jackiw and L. Susskind, {\it
Ann. of Phys.} {\bf 93} (1975) 267; S. Coleman, ibid. {\bf 101} (1976) 239.
\item{{\bf [13]}} J. Schwinger, {\it Phys. Rev.} {\bf 209}(1951) 749.
\item{{\bf [14]}} W. R. Guti\'errez, {\it Nucl. Phys.} {\bf
B176} (1980) 185.
\item{{\bf [15]}} K. Hornbostel, S. J. Brodsky and H. C. Pauli,
{\it Phys. Rev.} {\bf D41} (1990) 3814.
\item{{\bf [16]}} P. P. Srivastava, preprints CBPF-010/92 and CBPF-004/92.

\end